\documentclass[aps,prl,twocolumn,superscriptaddress]{revtex4}
\usepackage{multirow}
\usepackage{graphicx}
\usepackage{longtable}
\usepackage[utf8]{inputenc}
\usepackage{epstopdf}
\usepackage{longtable}
\usepackage{textcomp}
\usepackage{color}
\usepackage{amsbsy}
\usepackage{amsmath}
\usepackage{amsfonts}

\begin{document}

\title{Novel type of atomic-scale spin lattice at a surface and its emergent Hall effect}

\newcommand{\kiel}{Institute of Theoretical Physics and Astrophysics,
Christian-Albrechts University of Kiel, Leibnizstrasse 15, D-24098 Kiel, Germany}
\newcommand{\fz}{Peter Gr\"unberg Institut and Institute for Advanced Simulation,
Forschungszentrum J\"ulich and JARA, Germany}
\author{M. Hoffmann}
\email{hoffmann@theo-physik.uni-kiel.de}  
\affiliation{\kiel}
\author{J. Weischenberg}
\affiliation{\fz}
\author{B. Dup\'e}
\affiliation{\kiel}
\author{F. Freimuth}
\affiliation{\fz}
\author{P. Ferriani}
\affiliation{\kiel}
\author{Y. Mokrousov}
\affiliation{\fz}
\author{S. Heinze}
\affiliation{\kiel}

\date{\today}

\begin{abstract}
We predict the occurrence of a novel type of atomic-scale spin lattice in an Fe monolayer
on the Ir(001) surface. Based on density functional theory calculations we parametrize a
spin Hamiltonian and solve it numerically using Monte-Carlo simulations. We find the
stabilization of a three-dimensional spin structure arranged on a $(3 \times 3)$ lattice. 
Despite an almost vanishing total magnetization we predict the emergence of a large anomalous
Hall effect, to which there is a significant topological contribution purely due to the real
space spin texture at the surface. 
\end{abstract}

\maketitle

Localized stable spin textures such as skyrmions or chiral domain walls have attracted much
attention recently due to their unique topological and transport properties~\cite{Nag2013,
Ryu2013,Emo2013} and potential applications in spintronics~\cite{racetrack,Kis2011,Fer2013,
Samp2013}. A key ingredient for their occurrence is the Dzyaloshinskii-Moriya (DM)
interaction~\cite{Dzyaloshinskii,PhysRev.120.91}, which arises due to spin-orbit interaction
in systems with broken inversion symmetry, as in the bulk of non-centrosymmetric crystals or
at surfaces and interfaces. Hall effects play an important role in these systems. For instance,
the spin-orbit torque originating from the spin Hall effect drives the motion of chiral domain
walls in ultrathin films very efficiently and very high speeds have been reported~\cite{Ryu2013,
Emo2013}. The topological Hall effect, defined as the contribution to the Hall resistivity due
to chiral spin texture, serves as one of the main tools to pinpoint the skyrmion phase in the
phase diagram of bulk alloys such as MnSi or
FeGe~\cite{PhysRevLett.102.186602,PhysRevLett.108.267201,PhysRevLett.106.156603}. 

The topological Hall effect in complex large-scale magnetic structures is normally described
assuming the adiabatic viewpoint of infinitesimally slowly varying spin texture~\cite{Nag2013}.
For skyrmions, the topological Hall resistivity can be factorized into the product of an
emergent magnetic field, which is the direct consequence of the non-zero topological charge,
and the topological Hall coefficient $R^{\rm top}$, which can be determined from the electronic
structure of the ferromagnetic crystal~\cite{PhysRevLett.93.096806,PhysRevLett.102.186602}.
The validity of this picture has been demonstrated for large-scale skyrmions in bulk
Mn$_x$Fe$_{1-x}$Si alloys~\cite{PhysRevLett.112.186601}. On the other side of the length scale,
the chirality-driven contribution to the anomalous Hall effect (AHE) has been predicted and observed
in bulk strongly-frustrated correlated oxides and bulk antiferromagnets, which exhibit
non-collinear magnetic order on the scale of 1~nm~\cite{NComm2014,PhysRevLett.112.017205,
PhysRevB.80.100401,PhysRevB.82.104412,PhysRevLett.98.057203,Taguchi30032001,Nature2010,
PhysRevLett.87.116801}.

Chiral domain walls and skyrmions with an extent of down to 1~nm can also occur at transition-metal
interfaces and surfaces~\cite{Heide2009,PhysRevLett.108.197204,nphys2045,Science-2013-Romming-636-9}.
It has been demonstrated that higher-order exchange interactions can play a crucial role in such
systems. They enforce the atomic-scale skyrmion lattice observed for an Fe monolayer on the Ir(111)
surface~\cite{nphys2045} and a conical spin spiral phase for a Mn double layer on W(110)~\cite{2MnW110}.
However, to date, very little is known both experimentally and theoretically about the Hall
effects in such complex nanometer-scale spin textures at surfaces and interfaces.

Here, we present a model system for a novel type of atomic-scale spin lattice at a transition-metal
surface and study its transport properties. In contrast to systems explored so far for skyrmion 
spin textures the local exchange interaction is antiferromagnetic in this system. 
Based on density-functional theory (DFT) and Monte-Carlo simulations, we find a complex 
three-dimensional spin structure on a (3$\times$3) lattice for an Fe monolayer on Ir(001), 
which has been shown to grow pseudomorphically~\cite{PhysRevB.76.205418}. 
This spin texture with angles close to 120$^\circ$ between adjacent spins arises due to the coupling
of spin spirals stabilized by exchange and DM interaction via the four-spin interaction. Using DFT
we demonstrate that, despite very small total magnetization, this non-trivial (3$\times$3) spin
lattice gives rise to the AHE, essentially larger than that observed in non-collinear bulk
compounds~\cite{NComm2014,PhysRevLett.112.017205,
PhysRevB.80.100401,PhysRevB.82.104412,PhysRevLett.98.057203,Taguchi30032001,Nature2010,
PhysRevLett.87.116801}. We further show the existence in this system of a sizeable surface 
``topological'' contribution to the AHE and orbital magnetization, which originate purely in the 
non-trivial real-space spin distribution and do not rely on the presence 
of spin-orbit interaction.

Nanoscale spin textures at transition-metal interfaces~\cite{nphys2045,Dupe2014,Palotas2014,Ebert2014}
can be treated employing a Hamiltonian on the discrete atomic lattice 
\begin{align}\label{eq:efunct}
\begin{split}
H\ =\ - &\ \sum_{ij} J_{ij} (\mathbf{M}_i\cdot\mathbf{M}_j) 
- \ \sum_{ij} \mathbf{D}_{ij} \cdot (\mathbf{M}_i \times \mathbf{M}_j)\\
- &\ \sum_{ijkl} K_{ijkl} \left[\left(\mathbf{M}_i\cdot\mathbf{M}_j\right)\left(\mathbf{M}_k\cdot
\mathbf{M}_l\right) + \dots \right]\\
- &\ \sum_{ij} B_{ij} (\mathbf{M}_i \cdot \mathbf{M}_j)^2
+ \ \sum_i K_\perp (\ M_i^z)^2\
\end{split}
\end{align}
which describes the magnetic interactions between the magnetic moments $\mathbf{M}_i$ of atoms at
sites $\mathbf{R}_i$. For the Fe monolayer on Ir(001), we used DFT to obtain the parameters for
the exchange interaction ($J_{ij}$), the DM interaction ($\mathbf{D}_{ij}$), the four-spin
interaction ($K_{ijkl}$) and the biquadratic exchange ($B_{ij}$) as well as a uniaxial
magnetocrystalline anisotropy ($K_\perp$). We applied the projector augmented wave (PAW)
method~\cite{PhysRevB.50.17953} as implemented in the VASP code \cite{VASP4,VASP5}. Computational
details are given in the supplementary material.

\begin{figure}[t!]
\includegraphics[width=0.46\textwidth,clip]{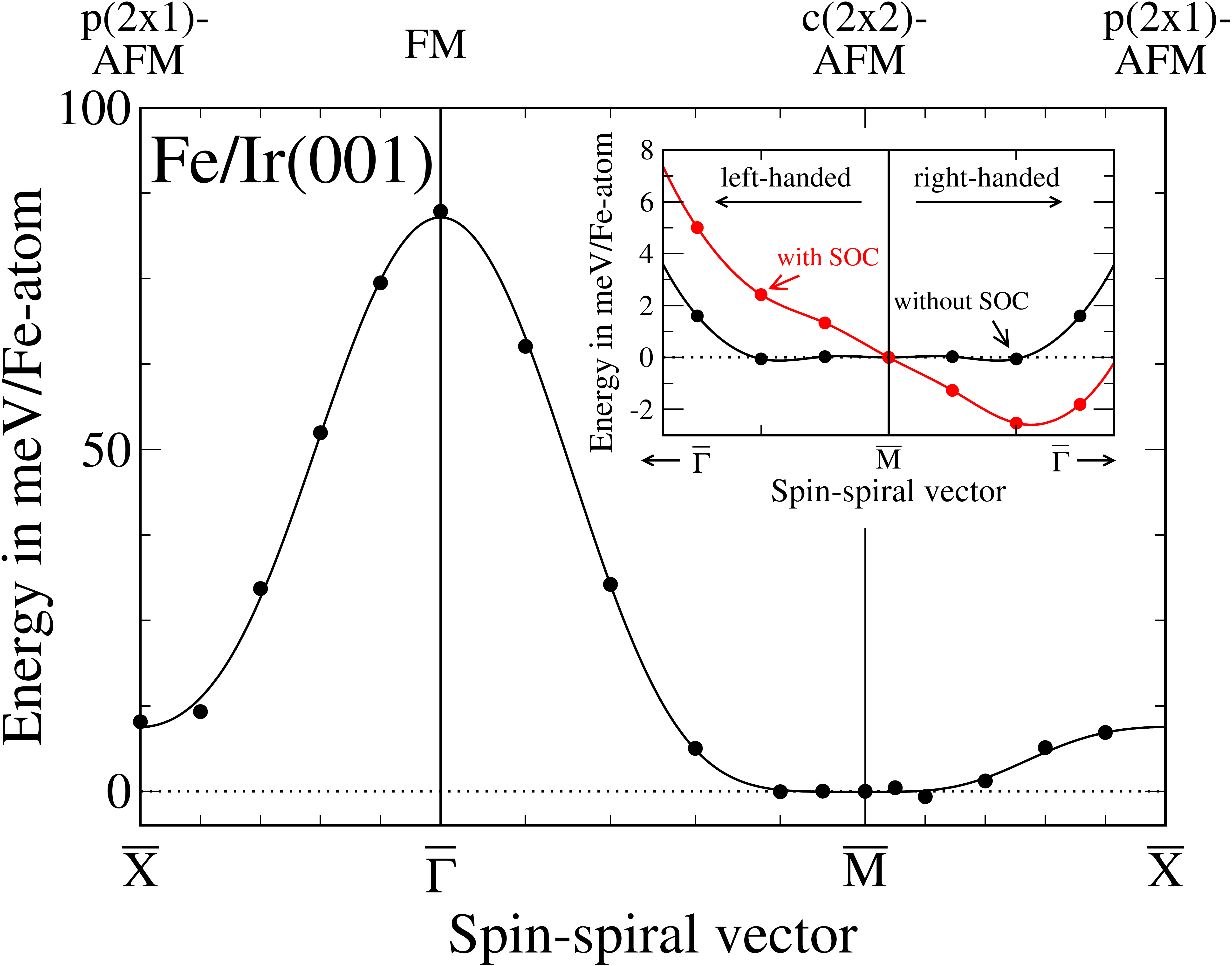}
\caption{
Energy dispersion of homogeneous flat spin spirals for Fe/Ir(001). The energies $E(\mathbf{q})$
(filled circles) are calculated via DFT along the high symmetry lines of the two-dimensional
Brillouin zone and given with respect to the $c(2\times 2)$ antiferromagnetic state. The solid
lines are fits to the Heisenberg model with up to sixth nearest-neighbors~\cite{note_J}.
The inset shows the energy dispersion close to the $\bar{{\rm M}}$ point for left- and
right-rotating spirals including SOC, i.e.~the effect of the DM interaction. 
}
\label{fig:exchange}
\end{figure}

To determine the exchange constants $J_{ij}$, we have considered flat spin spirals in which
the magnetic moments are confined in a plane with a constant angle between moments at adjacent
lattice sites propagating along high symmetry directions of the surface. Such a spin spiral
can be characterized by a wave vector $\mathbf{q}$ from the two-dimensional Brillouin zone
(BZ) and the magnetic moment of an atom at site $\mathbf{R}_i$, given by 
$\mathbf{M}_i=M (\sin{(\mathbf{q} \mathbf{R}_i}), \cos{(\mathbf{q} \mathbf{R}_i)},0)$ with 
the size of the magnetic moment $M$.

The calculated energy dispersion $E(\mathbf{q})$ of spin spirals for Fe/Ir(001) is displayed
in Fig.~\ref{fig:exchange}. At the high symmetry points we obtain collinear spin structures:
the ferromagnetic state at $\bar{\Gamma}$, the $c(2\times2)$ antiferromagnetic state at
$\bar{{\rm M}}$, and the $p(2\times1)$ antiferromagnetic state at $\bar{{\rm X}}$. Clearly,
the $c(2\times2)$ antiferromagnetic state is lowest in energy among the considered collinear
states in agreement with previous DFT studies~\cite{1306.5925v1}. The energy dispersion is very 
flat in the vicinity of the $\bar{{\rm M}}$-point due
to the frustration of exchange interactions. A fit to the Heisenberg model, i.e.~the first
term in Eq.~(\ref{eq:efunct}), with $J_{ij}$'s up to sixth nearest neighbors~\cite{note_J} 
leads to an excellent description as shown by the solid line in Fig.~\ref{fig:exchange} 
\cite{note_film}. 

Note, that the energy dispersion of Fe/Ir(001) is almost inverted with respect to Fe/Ir(111)
where the energy dispersion is flat around the $\bar{\Gamma}$-point, i.e.~the ferromagnetic
state \cite{PhysRevLett.96.167203,nphys2045}. Therefore, we can also expect complex
three-dimensional spin structures to occur here but of different type due to the nearest-neighbor
antiferromagnetic exchange.

By taking spin-orbit coupling (SOC) into account, we can determine the magnetocrystalline
anisotropy energy (MAE) defined as the energy difference between configurations with different
orientation of the magnetization. For the collinear state of lowest energy, i.e.~the
$c(2\times 2)$ antiferromagnetic state, we found an easy out-of-plane axis with a MAE of 
$K_\perp=-0.25$~meV. 

At a surface SOC also induces the DM interaction~\cite{Crep1998,nature05802}. In order to
determine its strength, we have calculated the total energy of a $120^\circ$ spin spiral along
the $\overline{\Gamma {\rm M}}$-direction in a $(3\times 1)$ super cell including SOC both
with a left-handed and a right-handed rotational sense. We find that spin spirals with a
right-handed rotational sense are lower by 7.3~meV/Fe atom. This energy difference allows
to calculate the value of the DM interaction within the nearest-neighbor approximation which
results in a value of $D_1=1.5$~meV. Including the DM interaction into the energy dispersion
of spin spirals leads to an energy minimum at an angle of about $138^\circ$ between adjacent
spins as shown in the inset of Fig.~\ref{fig:exchange}. 

From the energy dispersion of spin spirals, only the Heisenberg-type exchange can be obtained.
The impact of higher-order spin interactions can be determined by considering superposition
states of two spin spirals. If only Heisenberg-type exchange played a role all of these spin
states would be degenerate in energy. However, our DFT calculations show considerable energy
differences on the order of a few meV/Fe-atom (see supplementary material). From these
calculations, we determine that the nearest-neighbor four-spin, $K_{\rm4spin}$, and biquadratic,
$B$, interaction fulfill the condition $2K_{\rm4spin}+B=0.7$~meV.  

\begin{figure}[t!]
\includegraphics[width=0.30\textwidth,angle=270,clip=]{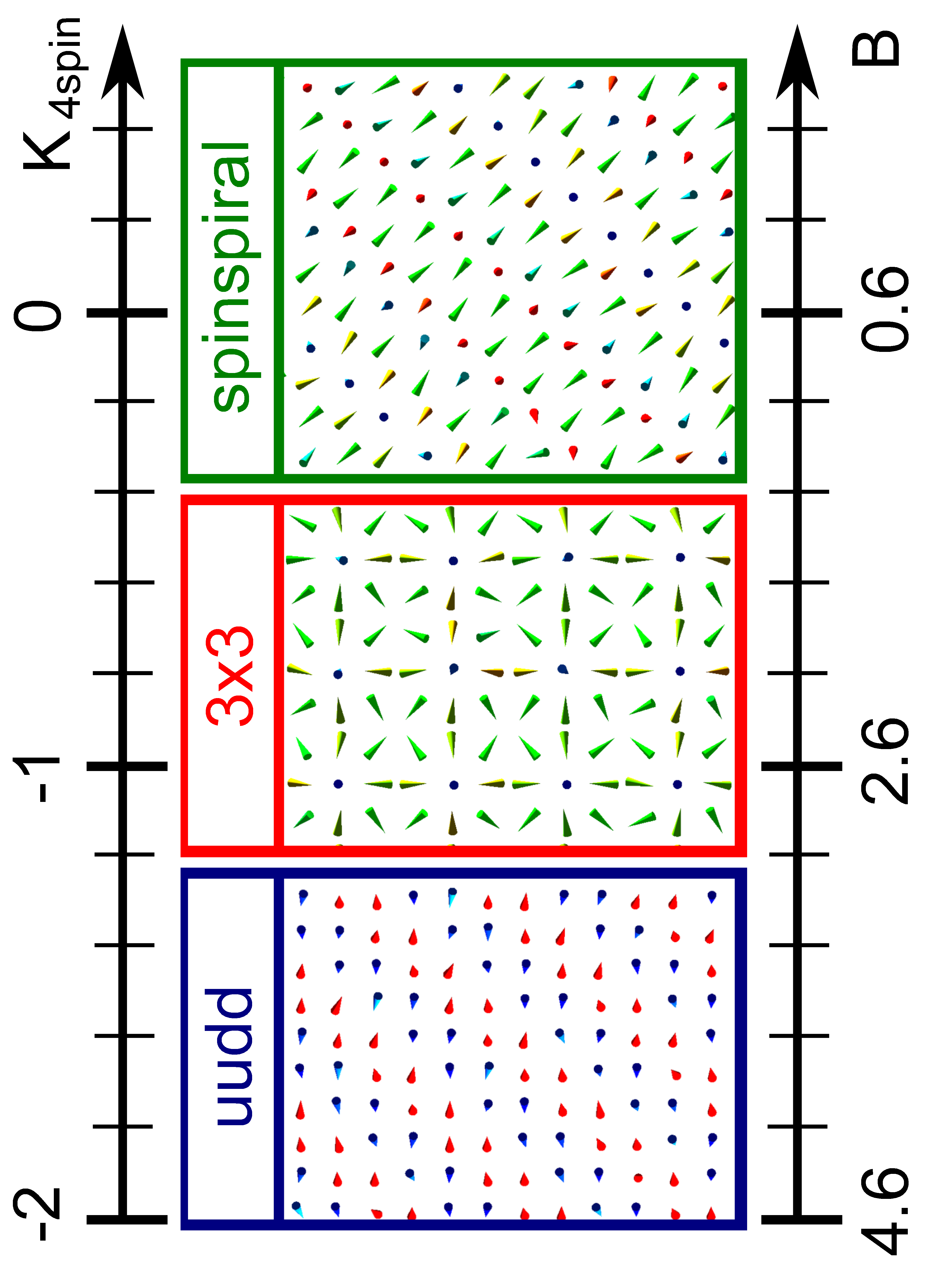} 
\caption{
Spin structures of lowest energy obtained via Monte-Carlo simulations as a function of the
strength of the four-spin interaction. The biquadratic interaction is changed according to
the condition $2K_{\rm 4spin}+B=0.7$~meV from the DFT calculations. The sketches only display
a small section of the actually simulated spin lattice.
}
\label{fig:mc}
\end{figure}

\begin{figure}[t!]
\includegraphics[width=0.35\textwidth,clip=]{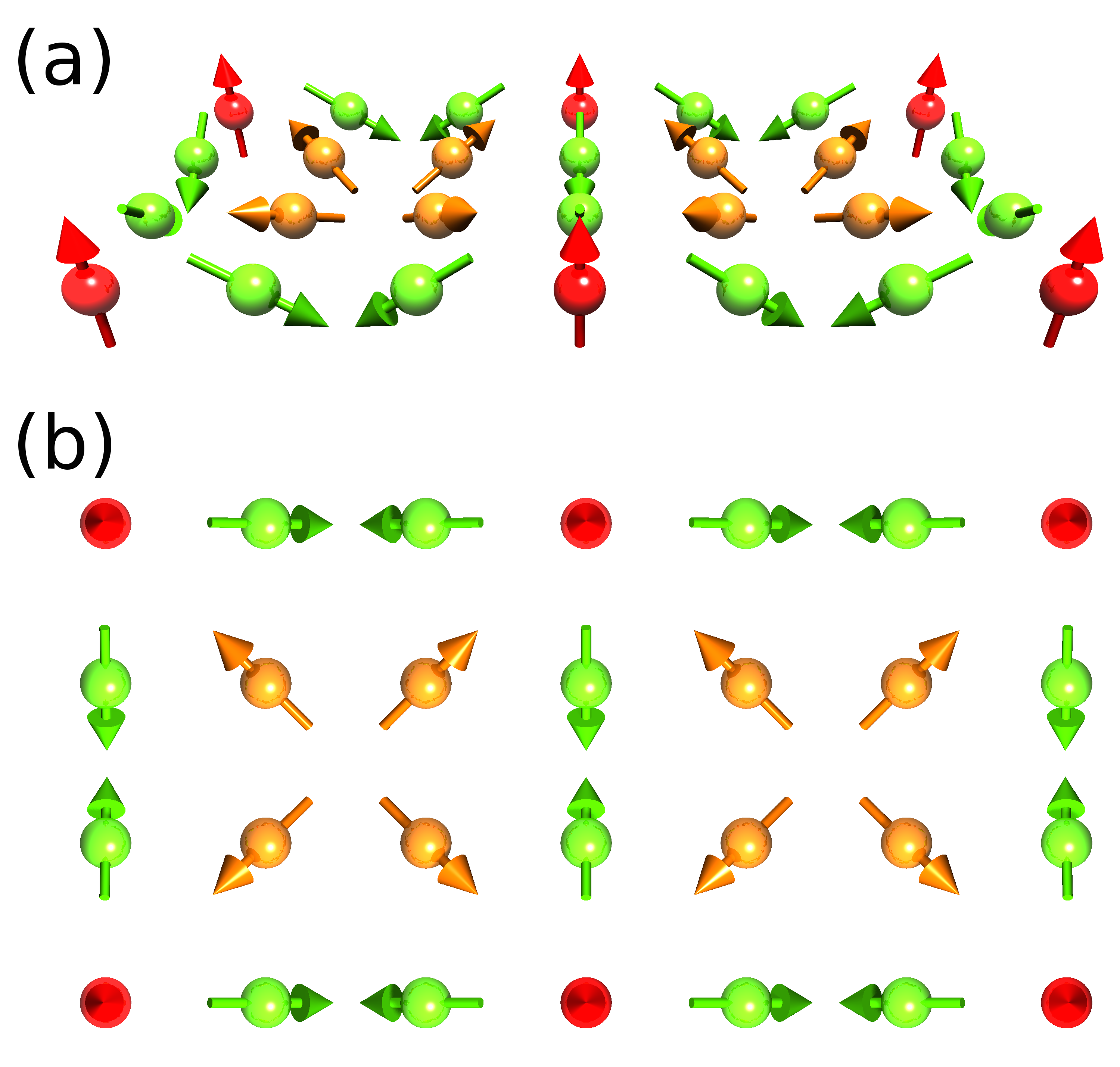}
\caption{
(a) Side view and (b) top view of the proposed atomic-scale (3$\times$3) spin lattice.
Two unit cells are shown.
}
\label{fig:3x3}
\end{figure}

The energy functional Eq.~(\ref{eq:efunct}) with the parameters from DFT can be minimized
using Monte-Carlo simulations based on the Metropolis algorithm. We have chosen a spin lattice
of $(66 \times 66)$ spins and used periodic boundary conditions. We have checked the impact of
the lattice size and of using open boundary conditions and found no effect on the obtained
ground state. In order to explore the impact of the higher-order spin interactions, which are
not univocally determined by our DFT calculations as discussed above, we have chosen different
values of $B$ and $K_{\rm 4spin}$ that are in accordance with the condition given above.  
We changed the value of the four-spin interaction in steps of 0.1 meV and the biquadratic
interaction and $J_3$ were modified accordingly~\cite{note_J}.

We found three different types of ground states depending on the value of $K_{\rm 4spin}$ as
shown in Fig.~\ref{fig:mc}. A large biquadratic interaction results in a so-called up-up-down-down
(\textit{uudd}) state since a collinear alignment of neighboring spins is preferred. However,
if the biquadratic interaction is reduced we find an atomic-scale non-collinear (3$\times$3)
spin lattice that is stabilized by the four-spin term. For a value of $K_{\rm 4spin}>-0.4$~meV
the four-spin term cannot couple the spin spirals and we obtain a spin spiral ground state with
an angle of about $140^\circ$ between adjacent spins. 

As shown in Fig.~\ref{fig:mc}, the novel (3$\times$3) spin lattice can occur for a large range
of values of the four-spin interaction. Its spin structure is shown in Fig.~\ref{fig:3x3}. The
spins at the corners of the unit cell point upwards perpendicular to the surface while the spins
along the sides rotate with an angle of $\approx 123^\circ$ from the surface normal. The four
spins in the center of the cell point towards the corners and with an angle of $\approx 22^\circ$
out of the film plane.  

The occurrence of this three-dimensional spin structure can be understood from the interplay
of the different interactions. The combination of exchange and DM interaction leads to a spin
spiral with an angle of approximately $120^\circ$ between adjacent spins and thus a periodicity
of 3 atoms (cf.~Fig.~\ref{fig:exchange}). For Fe biatomic chains on the $(5 \times 1)$
reconstructed Ir(001) surface, such a spin spiral state has been experimentally
observed~\cite{PhysRevLett.108.197204}. In the Fe monolayer on Ir(001), the four-spin interaction
can couple these spin spirals into a square lattice. Note, that there is an opposite rotational
sense of the spin rotation along the side and the diagonal of the unit cell. This results from
the antiferromagnetic exchange coupling between nearest neighbors which is stronger than the DM
interaction that would prefer a unique rotational sense along both directions. Due to this
peculiar competition of DMI and Heisenberg exchange, the spin lattice is extremely stable 
in an external magnetic field and cannot be destroyed up to 80~T as found in our MC simulations.
The transition temperature to the paramagnetic state is obtained at approximately 60~K.

\begin{figure}[t!]
\includegraphics[width=0.99\linewidth]{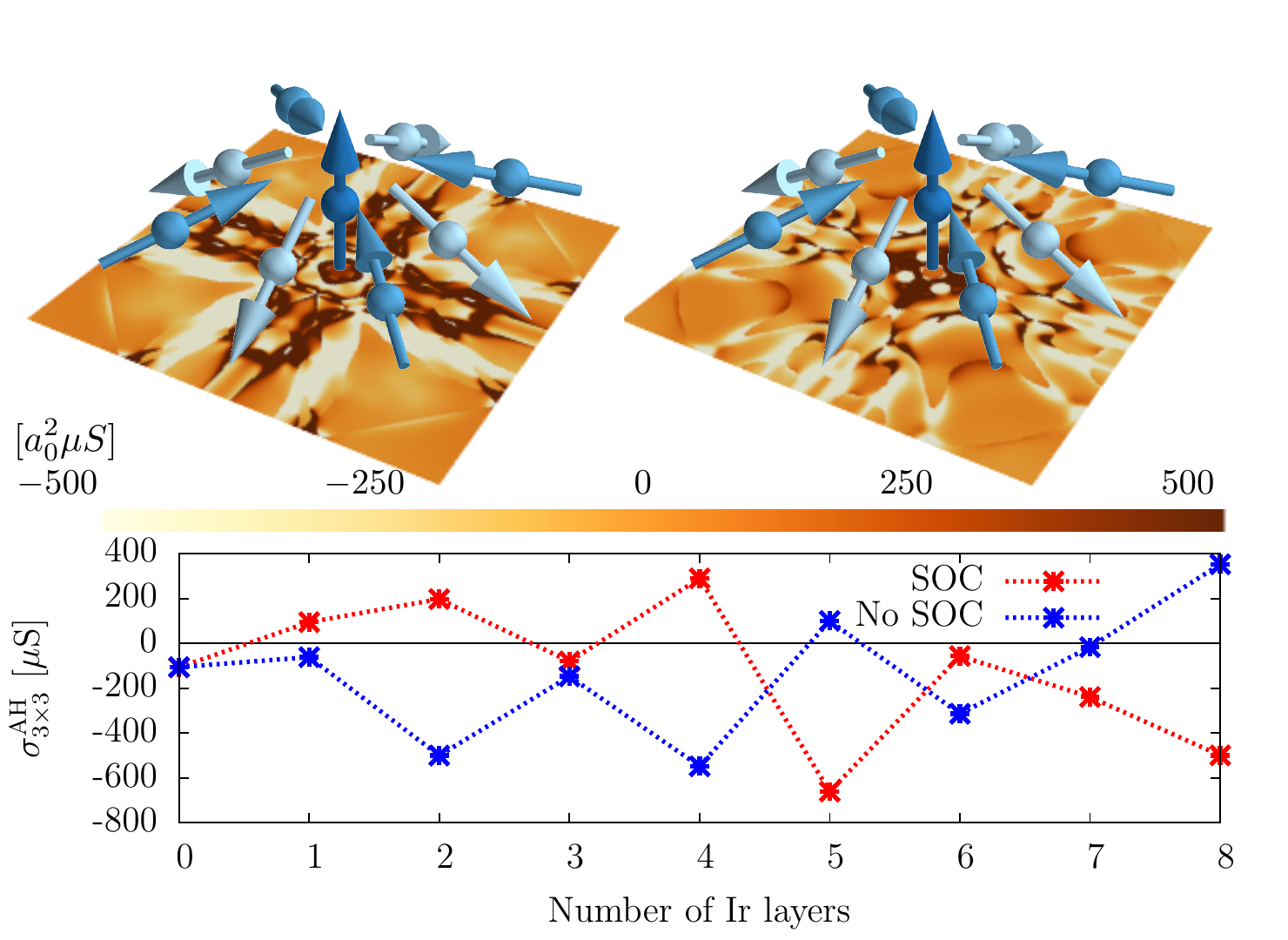}
\caption{
Top: BZ distribution of the Berry curvature without (left) and with (right) SOC for Fe monolayer
in ($3\times 3$) state with one layer of Ir substrate, superimposed with the real-space
distribution of the spins (blue arrows). Bottom:  calculated values of $\sigma^{\rm AH}_{3\times 3}$
as function of the Ir substrate thickness.
}
\label{fig:AHC_3x3}
\end{figure}

To investigate whether the (3$\times$3) spin texture results in non-trivial transport properties,
we compute from first principles~\cite{supplem} the intrinsic Berry curvature  contribution to
the $xy$-component of  the anomalous Hall conductivity (AHC) in the system
$\sigma^{\rm AH}_{3\times 3}=\frac{e^2\hbar}{(2\pi)^2}\int_{BZ}\Omega_{xy}(\mathbf{k})\,d\mathbf{k}$
~\cite{RevModPhys.82.1539}, where
\begin{equation}
\Omega_{xy}(\mathbf{k})=\sum_{n < E_F}\sum_{m\neq n}{\rm 2Im}
\frac{\langle\psi_{n\mathbf{k}}|v_x|\psi_{m\mathbf{k}}\rangle
   \langle\psi_{m\mathbf{k}}|v_y|\psi_{n\mathbf{k}}\rangle}
   {(\varepsilon_{m\mathbf{k}}-\varepsilon_{n\mathbf{k}})^2}
\end{equation}
is the Berry curvature of occupied states with $\psi_{n\mathbf{k}}$ as the Bloch states with
corresponding energies $\varepsilon_{n\mathbf{k}}$, and $v_i$ is the $i$'th Cartesian component
of the velocity operator. The results of our calculations for $\sigma^{\rm AH}_{3\times 3}$,
presented in Fig.~\ref{fig:AHC_3x3} as a function of the substrate thickness, indicate a
sizeable AHE in the (3$\times$3) spin lattice state with the magnitude similar to that of
bulk transition-metal ferromagnets~\cite{RevModPhys.82.1539,PhysRevB.89.014411,footnoteAHE}.
The large variation of the AHC with thickness, apparent from Fig.~\ref{fig:AHC_3x3}, is typical
for such effects as the AHE, spin Hall effect or the spin-orbit torque in the limit of ultrathin
films~\cite{PhysRevB.89.014411,PhysRevB.90.174423,PhysRevB.90.064406}.
 
In the context of thin magnetic layers of transition-metals on paramagnetic substrates, the
emergence of the large $\sigma^{\rm AH}_{3\times 3}$ appears rather surprising, since the
total magnetization of the system in the (3$\times$3) state is almost vanishing. By artificially
rotating the spin moments on the Fe atoms slightly away from their equilibrium directions we
acquire a complete suppression of the magnetization and observe that the values of 
$\sigma^{\rm AH}_{3\times 3}$ stay very close to those with small uncompensated magnetization.
This clearly distinguishes our case from the case of the AHE in collinear magnets, which relies
on non-vanishing macroscopic magnetization and presence of SOC~\cite{RevModPhys.82.1539}.

Another remarkable observation is that a large contribution to $\sigma^{\rm AH}_{3\times 3}$
is provided even without taking the SOC into account, as apparent from Fig.~\ref{fig:AHC_3x3},
where the values of the intrinsic AHC, computed with the SOC explicitly switched off in our
calculations, are presented in comparison with $\sigma^{\rm AH}_{3\times 3}$. Since the AHE
vanishes for any collinear magnetic state of our system without SOC, it allows us to define
the contribution to $\sigma^{\rm AH}_{3\times 3}$ without SOC as the ``topological" contribution
to the AHC, $\sigma^{\rm TH}_{3\times 3}$, which stems purely from the spin texture in real
space, and which does not rely on the presence of SOC. The particular symmetry of our system
which results in non-vanishing  $\sigma^{\rm TH}_{3\times 3}$, also gives rise to a finite
local scalar spin chirality $\mathbf{M}_i\cdot(\mathbf{M}_j\times\mathbf{M}_k)$, non-vanishing
when integrated over the unit cell~\cite{Taguchi30032001}. To distinguish our case from the
case of large two-dimensional skyrmions and bulk frustrated oxides, for which topological
contribution to the Hall effect in some cases can be described neglecting the spin-orbit
effects~\cite{PhysRevLett.93.096806,PhysRevLett.112.186601,PhysRevLett.98.057203,
Taguchi30032001,Nature2010}, for our class of systems we call the corresponding anomalous
Hall effect without SOC the {\it surface topological Hall effect}. Our calculations suggest
the existence of surface THE in transition-metal multilayers.  

Ultimately, the large values of $\sigma^{\rm TH}_{3\times 3}$ are due to a direct 
effect of the non-trivial real-space distribution of spin on reciprocal-space distribution
of the AHC, given by the Berry curvature~\cite{Fang03102003}. To convince ourselves in this
explicitly, we plot in Fig.~\ref{fig:AHC_3x3} the Brillouin zone distribution of the Berry
curvature computed with and without SOC for the system of an Fe layer in the ($3\times 3$)
spin state on one layer of the Ir substrate. As apparent from the case without SOC, there is
a very close correlation of the Berry curvature distribution with the spin-distribution in
real-space, while the effect of SOC is to provide an additional fine structure to this
distribution stemming from SOC-induced band splittings in the vicinity of the Fermi level.  
Thus the surface THE is more complex than the THE in large-scale skyrmions for which the
topological contribution to the THE $-$ the emergent field $-$ can be separated from the
electronic effects in a collinear host encoded in $R^{\rm top}$~\cite{PhysRevLett.102.186602,
PhysRevLett.112.186601}. The surface THE arises from a close intertwining of the real and
reciprocal space topology, which together play a role of a single multi-dimensional
topological object with non-trivial transport properties. 

The microscopic origin of the competition between non-collinearity and spin-orbit interaction 
for the AHE in such non-trivial surface spin textures as considered here, presents an
exciting direction to study both theoretically and experimentally~\cite{PhysRevB.86.245118,EV}. 
In particular, we conjecture that the surface THE is commonly an important part of the AHE
exhibited by complex spin structures at surfaces, such as nanoskyrmions~\cite{nphys2045}.
One of its prominent manifestations would be the contribution to the orbital magnetization at the
surface which is not originated in spin-orbit coupling~\cite{PhysRevLett.87.116801,
PhysRevLett.99.197202}. The orbital magnetization and the Hall effect have the same symmetry 
and indeed, our calculations reveal the formation of non-vanishing local atomic orbital moments 
at the surface of our system without spin-orbit. Without SOC, the magnitude of the maximal 
local orbital moment among the Fe atoms ranges from $-0.13$\,$\mu_B$ to $0.03$\,$\mu_B$ 
depending on the substrate thickness, which is similar in magnitude to that obtained with 
spin-orbit interaction included. Such ``topological" orbital magnetization could be readily
addressed experimentally by surface techniques.
 
B.D. and S.H. thank the Deutsche Forschungsgemeinschaft for financial support under project
DU1489/2-1. Y.M., F.F. and J.W. acknowledge funding under HGF Programme VH-NG 513 and DFG
SPP 1568. We gratefully acknowledge J\"ulich Supercomputing Centre, RWTH Aachen University
and HLRN for providing computational resources. It is our pleasure to thank Gustav Bihlmayer
for many insightful discussions.


\begin{thebibliography}{49}
\bibitem[{not({\natexlab{a}})}]{note_J}
\eprint{We have obtained the following values of the exchange constants $J_n$
  between $n$-th nearest neighbors in the Fe monolayer on Ir(001):
  $J_1...J_6=-9.44, -4.06,-1.59, -0.51, 0.01, 0.18$~meV. Note, that we can only
  determine the value $J_3+0.5 B= -1.59$ meV from the fitting where $B$ is the
  nearest-neighbor biquadratic exchange interaction.}
\bibitem[{not({\natexlab{b}})}]{note_film}
\eprint{We have checked the influence of the Ir film thickness on the energy
  dispersion of spin spirals up to 15 Ir layers and found convergence of the
  exchange parameters for 11 Ir layers.}
\bibitem[{foo()}]{footnoteAHE}
\eprint{The values of the AHC in units of S/cm can be obtained by dividing the
  values in Fig.~4 with the thickness of the film, and they lie in the range
  between several hundreds and several thousands S/cm, depending on the
  thickness. For example, in case of 3-layer thick Ir substrate (0.82~nm total
  thickness) the total AHC constitutes $-$98~S/cm.}
\end{thebibliography}
\end{document}